%% file: lni-paper-example-de.tex
\documentclass[english]{lni}
\usepackage{booktabs}

\usepackage{subcaption}
\usepackage{tikz}
\usetikzlibrary{spy,calc}
\usepackage{comment}
\usepackage{cases}
\usepackage{smartdiagram}
\usepackage{float} 
\usepackage{multirow} 
\usepackage{stfloats} 
\usepackage{array, makecell}

\begin{document}
\title[]{Foveated Path Tracing with Configurable Sampling and Block-Based Rendering}
 \author[1]{Bipul Mohanto}{bipul.mohanto@uni-rostock.de}{0000-0001-9913-0013}
 \author[1]{Sven Kluge}{sven.kluge@uni-rostock.de}{0009-0008-3670-7098}
 \author[2]{Martin Weier}{martin.weier@hs-rm.de}{0000-0001-9685-5238}
 \author[1]{Oliver Staadt}{oliver.staadt@uni-rostock.de}{0000-0002-3074-943X}%
 \affil[1]{University of Rostock\\Institute for Visual and Analytic Computing\\Albert-Einstein-Str. 22\\18059 Rostock\\Germany}
 \affil[2]{RheinMain University of Applied Sciences\\Applied Computer Science and Visual Computing\\65195 Wiesbaden\\Germany}
\maketitle

\begin{abstract}
Path tracing offers high-fidelity rendering but remains impractical for real-time applications due to slow convergence and noise. We present a dynamic foveated path tracing technique that leverages visual perception by reducing sampling towards peripheral regions. Our system achieves up to 25-fold performance gains on complex scenes at $4K$ resolution with minimal perceptual degradation. We validate its effectiveness using structured error maps across varying sampling rates and foveated region sizes, establishing a foundation for future research in perceptual photorealistic rendering.
\end{abstract}

\input{sections/introduction}

\input{sections/related_works}

\input{sections/background}
\input{sections/methodology}

\input{sections/result_and_discussion}
\input{sections/conclusion}
\input{sections/acknowledgment}

\printbibliography

\end{document}

%% file: sections/introduction.tex
\section{Introduction}
\label{_introduction}
\begin{figure}[t]
  \centering
  \includegraphics[width=1.0\textwidth]{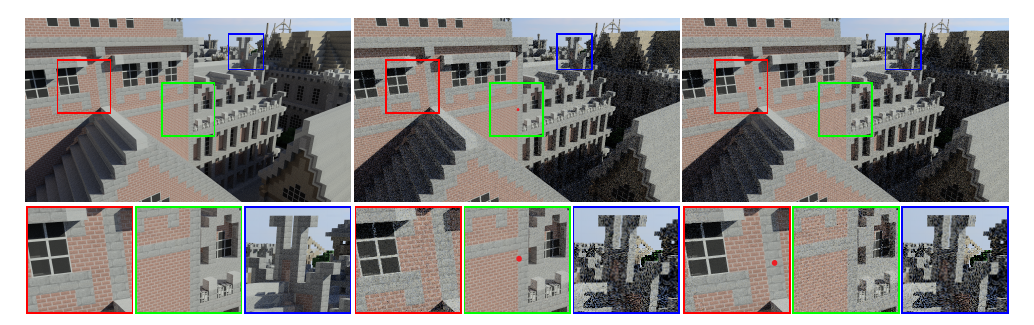}
  \caption{Foveated path tracing on the Rungholt scene (6.7M triangles, 4K). Left: Uniform 32 spp yields 0.37 fps. Middle/Right: Foveated sampling (32/16/8 spp) achieves 8.52 fps with more than $22\times$ speed-up. Fixations (red) are center and top-left.}
  \label{fig:teaser}
\end{figure}

Path tracing is a Monte Carlo-based rendering algorithm that accurately simulates global illumination by tracing the light paths as they interact with surfaces in a virtual scene. It enables photorealistic effects, such as soft shadows, caustics, diffuse interreflections, reflections, and refractions~\cite{kajiya1986}, which are challenging with traditional rasterization. Consequently, path tracing has become the \emph{de facto} standard in offline rendering for visual effects, architectural visualization, and product design. However, path tracing suffers from slow convergence~\cite{Roth2015}. Halving the noise requires quadrupling the sample count~\cite{Koskela2018}, making the method impractical for real-time applications on consumer-grade hardware.

While recent advances in GPU technology have opened the door to real-time path tracing research, most optimizations are tightly coupled to hardware architecture specifications. In contrast, foveated rendering offers a hardware-agnostic solution that exploits the non-uniform sensitivity of human vision to reduce rendering load without compromising perceived visual quality. It has traditionally utilized rasterization technology, which has been the predominant approach in real-time computer graphics. Nonetheless, there has been a gradual transition towards a physically accurate path tracer. Unlike rasterization, path tracing offers better flexibility for per-pixel manipulation in foveated rendering~\cite{Koskela2018}. Furthermore, Koskela et al.~\cite{koskela2016} demonstrated that approximately 94\% of rays may be omitted in foveated rendering, offering substantial potential for accelerating path tracing.

Although foveated rendering is particularly effective in single-view displays with a wide field of view (FoV), it is also applicable to display walls, single desktops, or even smartphones~\cite{kim2021efficient}. For instance, at a $60$ cm viewing distance, the human fovea subtends only $1.5^\circ-2^\circ$, covering roughly 3\% of a 21-inch display~\cite{duchowski2017eye}. The slow convergence of real-time further worsens with the increasing demand for higher pixel density, refresh rate, and FoV, especially in the immersive head-mounted displays (HMDs). Contemporary virtual reality (VR) headsets achieve angular resolutions around 35 cycles/degree and refresh rates near 90 Hz~\cite{Guanjun2018}, whereas the human fovea can resolve up to 60 cycles/degree~\cite{adam2019}. Achieving perceptual parity would require $6–10\times$ more pixels and $10–20\times$ faster refresh rates~\cite{Cuervo2018}, posing significant challenges for real-time path tracing.

In this study, we make the following contributions. We introduce a block-based foveated path tracing method that partitions the image into three perceptual regions: a high-fidelity foveal zone rendered per-pixel, and intermediate and peripheral zones rendered using configurable $(n\times n)$ and $(m\times m)$ pixel blocks, respectively. The sampling rates and block sizes are user-adjustable, enabling flexible control over the quality–performance trade-off. To address undersampling artifacts, missing pixels are reconstructed using linear interpolation. Compared to uniform sampling, our method achieves over $25\times$ performance improvement with minimal perceptual loss.

%% file: sections/related_works.tex
\section{Related Work}
\label{sec_related_work}
Foveated rendering has historically relied on the raster-based pipeline, which dominated the field for over three decades~\cite{MOHANTO2022}. However, recent advances in hardware-accelerated ray tracing have sparked a shift towards ray-based approaches. Fujita and Harada~\cite{fujita2014} were among the first to propose foveated sampling in ray tracing using a K-nearest neighbor filter. Nonetheless, the proposed method did not follow the accurate human contrast sensitivity function. Weier et al.~\cite{weier2016} introduced a temporal foveated ray tracing method, while Siekawa et al.~\cite{adam2019} implemented sparse sampling masks to reduce the number of primary rays based on gaze. Viitanen et al.~\cite{viitanen2018} and Peuhkurinen and Mikkonen~\cite{grapp21} explored sampling in log-polar space to reduce ray counts further. Yang et al.~\cite{yang2020} proposed variable-rate sampling, and Kim et al.~\cite{kim2021} introduced adaptive supersampling guided by perceptual importance. Dahlin and Sundstedt~\cite{dahlin2021} demonstrated a hybrid approach combining variable-rate shading with inline ray tracing, enabling perceptually plausible reflections at a fraction of the native resolution.

Foveated path tracing, a more recent and less explored extension of ray tracing, remains a challenging area of research. Koskela et al.~\cite{Koskela2018, Koskela2019} made notable advances by introducing foveated path tracing for HMDs. Their contributions include the first real-time foveated path tracing preview system~\cite{Koskela2018} and a log-polar ray mapping scheme for efficient sampling~\cite{Koskela2019}. Lotvonen et al.~\cite{Lotvonen2020} extended this work by applying machine learning techniques for foveated denoising. Kim and Sung~\cite{kim2021efficient} developed a lightweight foveated path tracing pipeline designed specifically for mobile VR platforms. Polychronakis et al.~\cite{Polychronakis2021} studied eccentricity-based sampling thresholds where foveated path tracing has been employed on the G-buffer. Compared to their study, which used post-process emulation, our method operates directly in the path tracing pipeline and supports real-time sampling. Similarly, Calla et al.~\cite{Calla2023} proposed a peripheral sample reducing technique. While their approach focused solely on shading reduction using per-pixel sampling, our technique jointly reduces both shading load and rendering resolution. This work is an expansion of the poster abstract presented at STAG 2023~\cite{mohanto2023}, incorporating improved environment lighting that elevates perceptual quality without incurring additional computational expense.


%% file: sections/background.tex
\section{Human Eye and Vision}
\label{_background}
Human vision is arguably the most complex of all biological senses that begins with the eye and ends at the visual cortex. The pupil allows light from the environment to enter the eye, which is then focused by the cornea and lens to the retina~\cite{goldstein2021sensation}. The retina is packed with photoreceptors like rods and cones. However, the non-uniform distribution does not allow humans to retain the identical level of perception throughout the entire FoV. Among the approximate 6 million cones, most of them are densely packed in the central foveal region, about $5.2^\circ$ from the visual axis~\cite{goldstein2021sensation}. Due to these dense cone cells, foveal vision is fine-grained with the highest visual acuity. Through the parafovea ($5.2^\circ-9^\circ$) and perifovea ($9^\circ-17^\circ$), cone density decreases rapidly, and rod density increases greatly~\cite{andersen2002}. Although there are around $20$ times more rods than the cones, the maximum density is close to $20^\circ$ from the visual axis~\cite{gustavosterberg1937}. Due to their luminance sensitivity, peripheral rods make the foveated rendering flicker-prone~\cite{Kwonoh2019}. 

\begin{figure*}[t]
    \centering 
    \includegraphics[width=1.0\textwidth]{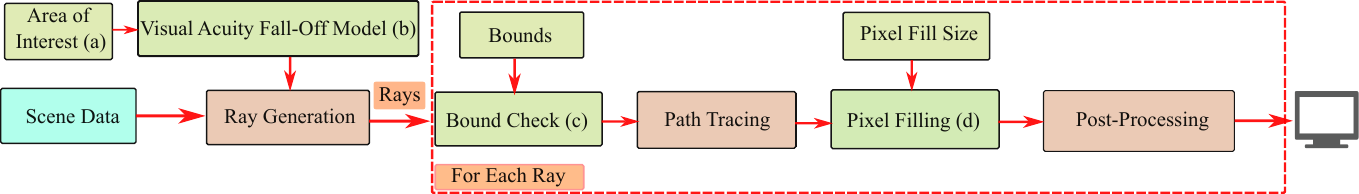}
    \caption{Overview of the proposed foveated path tracing pipeline. (a) The fixation defines the area of interest; (b) a visual acuity fall-off model determines sample distribution; (c) a bound check filters rays outside foveated regions; (d) missing pixels are reconstructed with pixel filling, followed by post-processing and final image display.}
    \label{fig:workflow_d}
\end{figure*}

%% file: sections/methodology.tex
\section{Methodology}
\label{_methodology}
We present a desktop-based foveated path tracing technique to enhance rendering performance for complex scenes. The proposed foveated pipeline (Figure~\ref{fig:workflow_d}) offers flexible control over performance and quality.

\subsection{Region Partitioning and Eccentricity Mapping}
At runtime, the image plane is partitioned into three regions: fovea, intermediate, and peripheral, centered around the gaze position (Figure~\ref{fig:workflow_d}. a). This multi-region structure strikes a balance between perceptual plausibility and computational cost, as finer subdivisions, while perceptually accurate~\cite{Geisler1998}, introduce additional overhead~\cite{Polychronakis2021}. The eccentricity angle $e$ (in degrees) is converted to a screen-space radius $r$ (in pixels) using a geometric mapping adapted from~\cite{matkovic1997}:

\begin{equation}
 r=\frac{2d_e\cdot R\cdot tan(e/2)}{S},   
\end{equation}

Here, $d_e$ is the viewing distance, $S$ is the physical display size, and $R$ is the resolution, all measured along the same axis. We employed a $4K$ monitor of size $(70.848\times39.852)\;cm^2$ placed at a standard $60\;cm$ viewing distance.

\subsection{Sample Distribution Model}
Sample allocation across regions follows a piecewise linear fall-off inspired by prior work~\cite{Guenter2012}, chosen for its implementation simplicity and tunable behavior. Due to the slow convergence, the large sample size is one of the main constraints of real-time path tracing. Conversely, the visual quality is highly sample-dependent. In previous studies, Weier et al., ~\cite{weier2016} quoted that more than 16 samples could be appropriate for real-time ray tracing. Likewise, Fujita and Harada~\cite{fujita2014} suggested 32 samples to avoid distracting visual artifacts. Therefore, we adopt a baseline of 32 spp in the foveal region, with progressively fewer samples in the outer zones:

\setcounter{equation}{2}
\begin{subnumcases}{f(e) = }
     \tau_1/px,  & $if,\; e\leq e_f$,   \label{eq_3a}  \\
     \tau_2/(n\times n)px,  & $else\; if, \; e_f < e \leq e_i$, \\
     \tau_3/(m\times m)px, & $otherwise.$
     \label{eq_3c}
\end{subnumcases}

Here, $\tau_1=32, \tau_2=16, \tau_3=8$ denote the sampling rates applied to foveal, intermediate, and peripheral regions, respectively. $e_f
$ is the foveated ($5.2^\circ$~\cite{Guenter2012, patney2016}) and $e_i$ is the intermediate ($17^\circ$~\cite{Roth2017}) visual angle. Pixel blocks of size $n=2$ and $m=4$ are used for downsampling in intermediate and peripheral zones, respectively. This configuration can be adjusted to explore quality and performance trade-offs. The three regions are seamlessly blended within a single render pass (Figure~\ref{fig:workflow_d}. b).

\subsection{Bound Check and Reconstruction}
Before ray generation, each screen-space pixel undergoes a binary region classification (Figure~\ref{fig:workflow_d}. c). Rays falling outside their designated region are discarded early. The remaining rays proceed through the full path tracing pipeline. Since fewer rays are cast than pixels in block-sampled regions, missing values are reconstructed via linear interpolation within each pixel block (Figure~\ref{fig:workflow_d}. d).

Kajiya-style path tracing~\cite{kajiya1986} is the backbone of our foveated path tracing. Additionally, we employed Disney Bidirectional Reflectance Distribution Function (BRDF)~\cite{burley2012physically}, multiple importance sampling~\cite{Veach1998}, and environmental light sampling. We limited the recursion depth to three bounces to avoid further computation load; however, we maintain full global illumination by supporting indirect bounces, albeit limited to three. Besides, we applied subpixel jitter, heuristic exposure compensation, and Reinhard tone mapping~\cite{salih2012} as post-processing for a mellow output. Our implementation supports temporal accumulation. Nevertheless, there is little scope for temporal accumulation in real-time applications, e.g., VR rendering with frequent eye movements. Therefore, we presented the scenes (Figure~\ref{fig:teaser},~\ref{fig:scenes}) in this study without temporal accumulation.

\subsection*{Software and Apparatus}
\label{_hardware}
The implementation is based on the OptiX 7.5 Graphics API with CUDA 12.1. The rendering computations were performed on an Intel Core i9-12900KS CPU and 128 GB of DDR5 system memory, NVidia GeForce RTX 3090 GPU featuring 24 GB of VRAM, and Studio Driver 536.99. For visual inspection and evaluation, we employed a Dell G3223Q 32-inch 4K UHD display, offering a pixel density of 54 pixels/cm and a nominal field of view of approximately $(175\times175)^\circ$.

%% file: sections/result_and_discussion.tex
\section{Results and Discussion}
\label{_results}
\begin{figure*}[t]
    \center
    \includegraphics[width=1.0\textwidth]{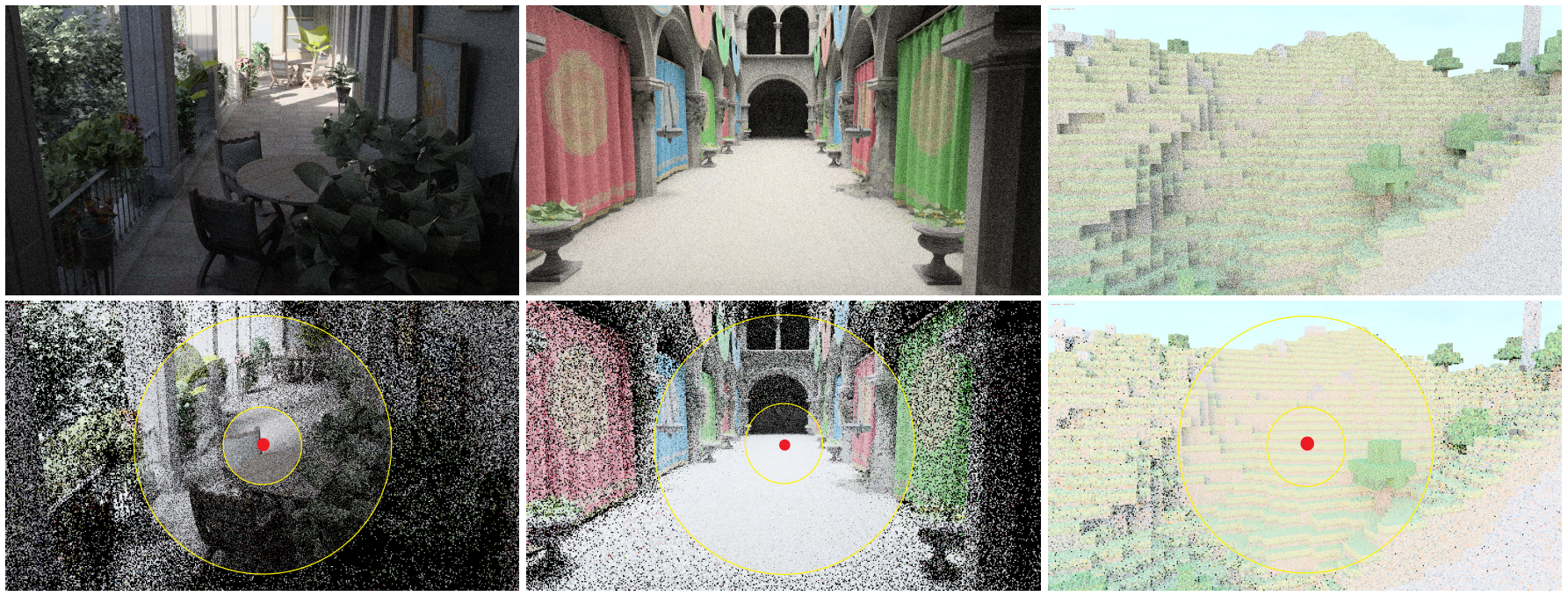}
    \caption{
    Visual comparison of uniform (top) and foveated (bottom) path tracing. The uniform mode uses 32 spp/pixel, while the foveated mode applies 32 spp/pixel in the fovea, 16 spp/($2\times2$) pixel block in the intermediate, and 8 spp/($4\times4$) pixel block in the periphery. Fixation points are marked in red. Scenes: San Miguel (left), Crytek Sponza (center), and Lost Empire (right).
    }
    \label{fig:scenes}
\end{figure*}

\begin{figure}[h]
    \centering
    \includegraphics[width=0.8\linewidth]{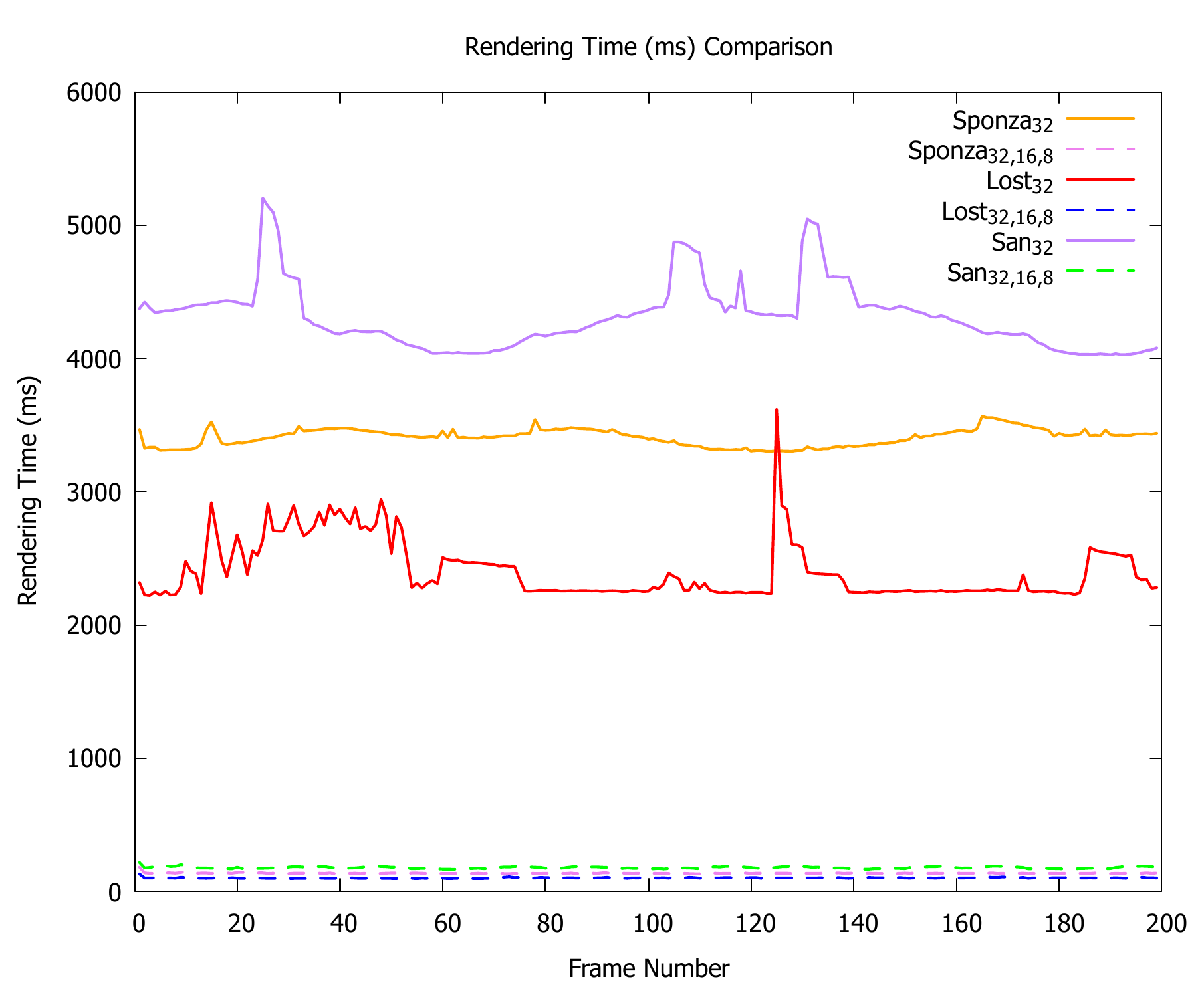}
    \caption{Rendering time for each scene, Crytek Sponza (Sponza), Lost Empire (Lost), and San Miguel (San). Solid lines denote uniform sampling (32 spp); dashed lines indicate foveated sampling (32/16/8).}
    \label{fig:fps_comp}
\end{figure}

We evaluated our dynamic foveated path tracing framework on three geometry-intensive scenes: Crytek Sponza, Lost Empire, and San Miguel (Figure~\ref{fig:scenes}). Although our system currently lacks a physical eye tracker, we emulate gaze behavior using predefined pseudo-fixations. Table~\ref{tab:1} summarizes the triangle count, average frame rates (fps), and speed-up achieved with uniform and foveated sampling configurations, measured on the hardware setup described above. The rendering time comparisons are further visualized in Figure~\ref{fig:fps_comp}.

\begin{table}[h]
   \centering
    \begin{tabular}{p{1.9cm}p{1.2cm}p{2.8cm}p{2.8cm}p{1.7cm}}
    \toprule
\textbf{Scene} & \textbf{Triangles} & \textbf{average fps (uniform)} & \textbf{average fps (foveated)} & \textbf{Speed-up ($\times$)} \\ \midrule
Lost Empire & 224,998 & 0.42 & 10.20 & $23.29$ \\ 
Crytek Sponza   & 262,267 & 0.24 & 7.48 & $30.17$\\ 
San Miguel  &  9,980,699 & 0.23  & 5.71 & $23.83$ \\ \bottomrule 
\end{tabular}
\caption{\label{tab:1} Average rendering performance across three scenes. Columns report triangle counts, frame rates under uniform (32 spp) and foveated sampling (32/16/8), and overall speed-up.}
\end{table}

Without foveation, the scenes achieve average frame rates below 0.5 fps, leads to non-interactive frame rates at $4K$ resolution. Our foveated approach yields substantial improvements: over 10 fps for Lost Empire, 7.5 fps for Crytek Sponza, and 5.7 fps for San Miguel, corresponding to an average speed-up of more than $25\times$. As shown in Figure~\ref{fig:fps_comp}, rendering time scales proportionally with scene complexity, but is consistently reduced under foveated sampling.

Visual outputs are illustrated in Figure~\ref{fig:scenes}. As expected, due to lower sample counts, the foveated results exhibit increased noise in the peripheral regions. However, the resulting artifacts are largely imperceptible during fixation.

\subsection{Effect of Sampling Strategy}
To further investigate the trade-off between sample count and performance, we evaluated multiple sampling configurations on the Crytek Sponza scene. As shown in Table~\ref{tab:diff_sample_num_sponza}, reducing sample counts in the intermediate and peripheral regions significantly improves frame rates. The most aggressive configuration (32/2/1 spp) achieves a $56\times$ speed-up over uniform rendering, with frame rates exceeding 17 fps. While performance improves linearly with reduced sample counts, aggressive foveation leads to perceptual artifacts in the periphery. Therefore, an optimal configuration must consider the task, scene complexity, and perceptual thresholds.

\begin{table}[h]
\centering
\begin{tabular}{c|ccc|cc}
\hline
\multicolumn{1}{l}{}           & \multicolumn{3}{|c|}{\textbf{Different Sample Configurations}} & \multicolumn{1}{l}{} & \multicolumn{1}{l}{} \\ \cline{2-4}
\textbf{Scene}                          & \textbf{Fovea}        & \textbf{Intermediate}       & \textbf{Periphery}       & \textbf{average fps}                  & \textbf{Speed-up ($\times$)}             \\ \hline
\multirow{5}{*}{\makecell{Crytek Sponza}} & 32         &
 --        & --        & 0.30                 & --                   \\  
                               & 32         & 16        & 8         & 7.48                 & 23.93                \\
                               & 32         & 8         & 4         & 10.90                & 35.33                \\
                               & 32         & 2         & 1         & 17.14                & 56.13                \\ \cline{1-6} 
\end{tabular}
\caption{\label{tab:diff_sample_num_sponza} Rendering performance for Crytek Sponza under different sampling configurations in the intermediate and Periphery. }
\end{table}

\begin{figure*}[t]
    \centering
    \includegraphics[width=1.0\linewidth]{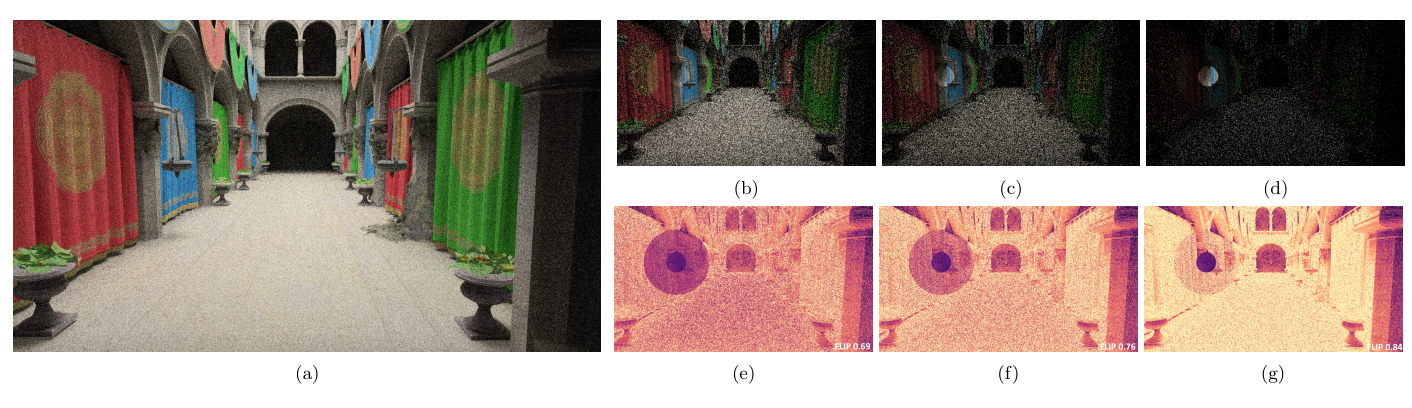}
    \caption{\label{fig:samp_vs_rendering_time}
    \reflectbox{F}LIP perceptual error maps ((e)-(g)) for different sample combinations. Figure (a) is the reference, (b)-(d) are the sample combinations mentioned from Table~\ref{tab:diff_sample_num_sponza}. Black = no error; yellow = increased perceptual deviation.}
\end{figure*}

To assess visual degradation, we employed the \reflectbox{F}LIP metric~\cite{Andersson2020}, which approximates perceptual differences between the rendered and reference image. Figure~\ref{fig:samp_vs_rendering_time} presents \reflectbox{F}LIP error maps for various sampling strategies. \reflectbox{F}LIP values close to zero in foveal regions indicate perceptual fidelity, while increasing values in peripheral areas reflect controlled degradation under low-sample conditions.

\begin{figure}[h]
    \centering
    \includegraphics[width=0.8\linewidth]{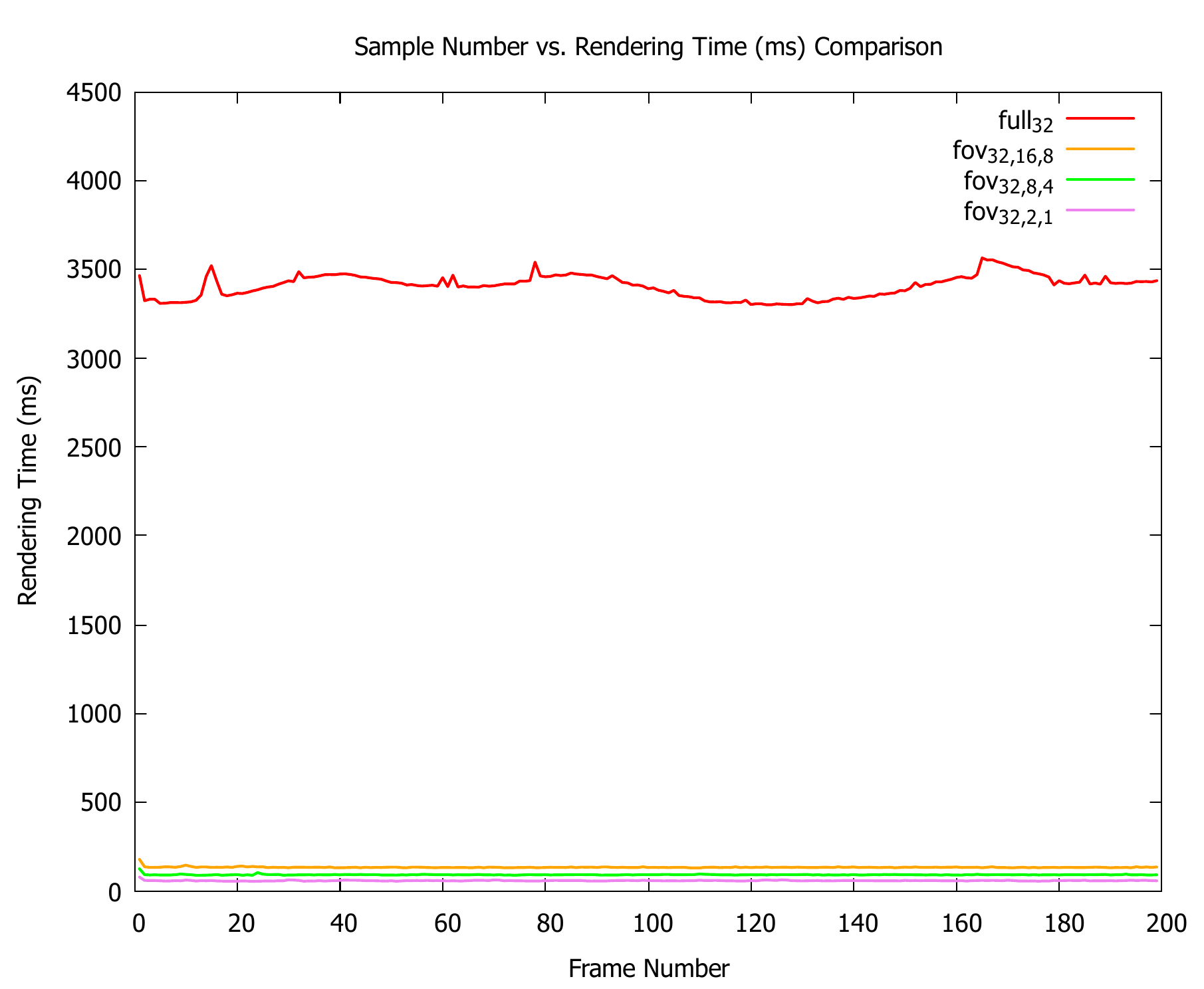}
    \caption{Rendering time for different sampling combinations on Crytek Sponza scene. The sampling combinations are in Table~\ref{tab:diff_sample_num_sponza}.}
    \label{fig:diff_sample_num_sponza}
\end{figure}

\subsection{Limitations}
This study presents several avenues for future improvement. First, due to the absence of a physical eye tracker, we relied on pseudo-gaze input; however, the system is readily compatible with a desktop-based eye tracker. Second, under dynamic viewing conditions, linear interpolation may fail to fully reconstruct missing pixels. We plan to explore more robust reconstruction techniques. Additionally, no denoising was applied in this work-integrating modern real-time denoisers is expected to further enhance perceptual quality. Building on these improvements, we aim to conduct a user study to determine perceptual thresholds for sampling versus visual fidelity.

%% file: sections/conclusion.tex
\section{Conclusion and Future Work}
\label{_conclusion}
In this work, we presented a dynamic foveated path tracing technique that combines perceptually guided sampling with block-based rendering to accelerate performance on pixel-intensive displays. By allocating samples based on visual eccentricity and employing configurable block sizes in non-foveal regions, our method achieves over $25\times$ speed-up compared to uniform sampling at $4K$ resolution. These results demonstrate the effectiveness of foveated path tracing as a practical and scalable solution for real-time photorealistic rendering.

While effective, the current implementation has limitations. Integration with a real-time eye tracker and denoiser is anticipated to further enhance graphics fidelity. The bound check discards rays outside circular regions post-generation, leading to inefficiencies due to GPU workload subdivision. Future work will focus on optimizing ray generation to avoid off-bound rays entirely. We also aim to extend the current piecewise linear acuity fall-off model by an alternative quadric fall-off to refine sample allocation. Based on these observations, future work will also include a user study to evaluate perceptual thresholds and guide adaptive sampling strategies for foveated path tracing.

%% file: sections/acknowledgment.tex
\section*{Acknowledgments}
\label{_acknowledgements}
The authors would like to thank the 3D scene creators, Frank Meinl (Crytek Sponza), Morgan McGuire (Lost Empire), Guillermo M. Leal Llaguno (San Miguel 2.0), Kescha (Rungholt), and Poly Heaven for generously providing their HDRI lighting assets.